# Unveiling the Magmatic Architecture Beneath Oceanus Procellarum: Insights from GRAIL Mission Data

Meixia Geng[1], Qingjie Yang[1], Chaouki Kasmi[1], J. Kim Welford[2], Alexander L. Peace[3]

[1]Directed Energy Research Centre, Technology Innovation Institute, Abu Dhabi, UAE.

[2]Memorial University of Newfoundland, Department of Earth Sciences, St. John's, Newfoundland and Labrador, Canada.

[3]School of Earth, Environment and Society, McMaster University, Hamilton, ON L8S 4K1, Canada.

## Abstract

The Oceanus Procellarum region, characterized by its vast basaltic plains and pronounced volcanic activity, serves as a focal point for understanding the volcanic history of the Moon. Leveraging the Gravity Recovery and Interior Laboratory (GRAIL) mission data, we imaged the magmatic structures beneath the Oceanus Procellarum region. Our 3D density models uncover pronounced linear magmatic structures along the Procellarum's western border and significant intrusions within the northern and southern Marius Hills. Crucially, they reveal three narrow near-horizontal sheeted magmatic structures, 80-150 km long, extending from near-surface to 6-7 km depth, which we identified as sill-like magmatic conduits. These magmatic conduits connect the Marius Hills' northern and southern intrusions and bridge them with the Procellarum's western border structures. These discoveries suggest that sill-like magmatic conduits likely serve as central pathways facilitating magma transport across various volcanic systems and furthermore indicate widespread magmatic connectivity beneath the Oceanus Procellarum.

# Introduction

Oceanus Procellarum, the largest mare on the Moon's near side, is characterized by extensive basaltic plains and significant volcanic activity[1]. Uncovering its subsurface structures is pivotal for understanding lunar volcanism. Gravity data, which have played a crucial role in studying magma plumbing systems on rocky planets and satellites[2, 3, 4, 5, 6], provide the ideal tool for such an investigation of Oceanus Procellarum. Forward modeling and inversion of gravity anomaly data enable the creation of density models of various magmatic structures that exhibit a density contrast with the surrounding rocks, such as shallow intrusions[7, 8, 9], magmatic-related deposits[10], and plutons[11, 12]. Data from NASA's Gravity Recovery and Interior Laboratory (GRAIL) mission have generated a high-resolution global map of the Moon's gravity field. By modeling the gravity field to a spherical harmonic degree and order of 1200, a spatial resolution of 4.5 km by 4.5 km at the Moon's equator was achieved[13, 14]. Combined with topographic data from the Lunar Orbiter Laser Altimeter (LOLA), these data aid the detection of small-scale features such as dykes[7], ancient igneous intrusions[3], buried impact craters[15], and empty lava tubes[16, 17].

The volcanic legacy of the Oceanus Procellarum is exemplified by the Marius Hills volcanic complex (MHVC), a 35,000 km$^2$ plateau, which exhibits the densest concentration of extrusive volcanic features on the Moon in the form of volcanic domes/shields and cones[18, 19]. This region has consistently emerged as a prominent site for lunar exploration, primarily due to its distinctive geological characteristics and the valuable insights it offers into lunar volcanism[20, 21, 22, 23]. Utilizing gravity data from the Lunar Prospector spacecraft[24] and the GRAIL mission[25, 26], two quasi-circular, positive gravity anomalies within the MHVC (north and south) were identified and modeled as two massive sill/dyke groups that are responsible for the volcanic activity in this region. Investigation into the surface distribution of volcanic vents confirms the presence of the zones of magma accumulation[27]. However, magmatic structures beneath the MHVC and surrounding areas and processes by which magma is transported to the shallow crustal depths remain enigmatic to date[28, 29].

GRAIL gravity data revealed a pattern of narrow linear anomalies that border the Procellarum KREEP Terrane (PKT)[5]. The source of anomalies is interpreted to be solidified lava within rifts and the feeder dykes beneath them, which likely served as the magma plumbing system for much of the nearside mare volcanism. The MHVC is positioned roughly 200 km east of the PKT's western border structures. Previous research has primarily examined these volcanic systems independently[5, 24, 25, 26, 27, 30], overlooking the critical magmatic structures between them and their collective impact on volcanism in the Oceanus Procellarum. This oversight underscores a significant gap in our understanding of the magmatic architecture beneath the MHVC and the PKT's western border, highlighting the importance of considering these systems as interconnected entities to comprehend the extensive volcanic activity fully in the Oceanus Procellarum region.

To better understand the magmatic architecture beneath the Oceanus Procellarum, we apply a three-dimensional (3D) inversion technique to the GRAIL gravity data to construct detailed 3D density models. The magmatic structures, characterized by higher density than the host rocks, can be recovered through the inversion of the Bouguer gravity anomaly, which plays a key role in understanding the magmatic sources and processes[2, 7, 9, 31]. This study encompasses the southern and northern MHVC, the western border structures of the PKT, and their neighboring areas, enabling a comprehensive understanding of their interconnections, and facilitating their integration into the wider PKT region. The detailed imaging of these magmatic structures provides valuable information regarding the formation of volcanic landforms, magma transport, and energy exchange across various lunar systems.

## Results

### Bouguer gravity data and topography

Fig. 1a shows the topography of the lunar nearside derived from the MoonTopo2600p model collected by the Lunar Reconnaissance Orbiter mission[32]. GRAIL gravity data extracted from the GRGM1200B model[13] were used for this study. The Bouguer anomaly (Fig. 1b) used for imaging the subsurface structures was derived from SHTools[33] with a reference radius of 1,738 km. To remove the gravitational contribution of the surface topography, and a crust density of 2,560 kg/m³ were used for Bouguer correction[34]. A higher density of 2,800 kg/m³ was also tested to derive the Bouguer anomaly, which was then inverted to construct density models. As illustrated in Supplementary Fig. 1, the density used for Bouguer correction has a minimal impact on the resulting structures of models, which are presented in terms of density contrasts. It does, however, influence the actual density values used in their interpretation. In the inversion density models, the average density of the crust is calculated as 2,588 kg/m³ with a Bouguer correction of 2,560 kg/m³, and it increases to 2,827 kg/m³ when the Bouguer correction is adjusted to 2,800 kg/m³. This calculated average density of 2,588 kg/m³ is in good agreement with laboratory measurements from lunar feldspathic meteorites and rocks collected during the Apollo missions[35], which have an average bulk density reliably estimated at 2,580 ± 170 kg/m³. To eliminate long-wavelength variations in the crustal structure and short-wavelength noise, we filtered the Bouguer gravity data to include data between degree and order 7 and degree and order 660, yielding a half-wavelength surface resolution of ~8 km. A cosine taper was implemented between degrees 550 and 660 to reduce high-frequency ringing[36].

Study area A (Fig. 1) covers MHVC and partial western border structures of the PKT[5]. The Bouguer gravity anomaly (Figs. 1b, 2b) reveals two quasi-circular mass anomalies, marked as Marius-NH (northern half) and Marius-SH (southern half) based on their locations relative to the Marius crater[26]. The two anomalies exhibit maximum amplitudes of 176 mGal and 167 mGal, respectively. Marius Hills plateau, corresponding to the northern anomaly, contains most domes and cones situated in a localized highland. In contrast, the southern anomaly covers fewer volcanic domes without elevated topography. A narrow linear positive gravity anomaly, ranging from ~40 to ~70 mGal, connects these two quasi-circular anomalies, forming a "bridge". Furthermore, a significant high-gravity anomaly belt, spanning approximately 800 km with a NNW orientation and peaking at 175 mGal, is identified. This belt has been interpreted to be part of the magmatic border structures of the PKT region[5]. The linear gravity anomaly continues to extend NNW and diverges into three thinner branches at the northwestern border of the PKT (Figs. 1b, 3b). Study area B (Figs. 1, 3a, 3b) includes the western and middle branches of the northwestern border, each marked by distinct linear gravity anomalies with a general NNW orientation, spanning over 500 km and displaying values from 40 to 125 mGal. The western edges of the belt gravity anomalies (Fig. 2b, 3b) are generally coincident with steep topographical gradients (Figs. 2a, 3a), while their eastern sides show less topographical correlation. Prominent wrinkle ridges run parallel to the belt gravity anomaly in study area A (Figs. 2a, 2b, Supplementary Fig. 2a) and the middle branch gravity anomaly in study area B (Figs. 3a, 3b, Supplementary Fig. 2b), exhibiting an NNW orientation. Also, a distinct set of wrinkle ridges (Figs. 2a, 2b, Supplementary Fig. 2a) traversing the Marius Hills region predominantly adheres to this NNW trend[37].

**3D inversion density models from Bouguer gravity data**

Our 3D density models recovered from the Bouguer gravity data with a half-wavelength surface resolution of ~8 km facilitate the investigation of complex subsurface magmatic structures in detail. Figures. 2c-2e and 3c-3e display the 3D density models for the two study areas. The red surfaces indicate the density iso-surface of 2,720 kg/m$^3$, which outlines high-density bodies and are interpreted as magmatic bodies. The yellow surfaces depict a higher density iso-surface of 2,800 kg/m$^3$ to delineate even denser magmatic bodies. The top depth corresponds to either the shallowest limit of the magmatic bodies or the deepest extent of impact brecciation, where the density contrast fades. The bottom depth denotes the base of the magmatic bodies or where the density contrast with surrounding rocks significantly diminishes. Figures 4 and 5 show horizontal and vertical cross-sections through the inverted density models for the two study areas. The blue dashed lines indicate the 2,720 kg/m$^3$ density contour lines. One limitation and source of potential errors in interpretation may arise from the assumed density iso-surface of 2,720 kg/m$^3$ for the magmatic body, which does not consider potential density changes amongst the magmatic rocks and could over- or underestimate the size and depth of the magmatic bodies.

Figures 2c, 3c, 4, and 5 exhibit distinct belt structures oriented in the NNW direction with densities ranging from ~2,720 to ~3,200 kg/m³. These belt structures correspond to the belt gravity anomalies depicted in Figures 2b and 3b and match the PKT border magmatic intrusions as identified in the previous study[5]. The density model for area A (Figs. 2c, 4) shows that the border intrusion spans from near the surface down to depths of 10 to 13 km with a length of ~800 km, a thickness of 8 to 11 km, and a width of 130 km to 270 km. In study area B, the western branch intrusion is ~600 km in length, ~140 km in width, and ~7.5 km in thickness, and the middle branch intrusion is about 550 km long, 80 km wide, with a thickness ranging from 5 to 7.5 km. The western PKT border intrusions become thinner and narrower, and the depth of the top surface increases as the range goes north.

The inversion density model indicates that the Marius-NH intrusion (Fig. 2d) and Marius-SH intrusion (Fig. 2e), reported in literatures[24, 25], extend down to 11.5 km and 12.5 km beneath the surface, respectively, with the Marius-NH body spanning 105 km in diameter and the Marius-SH body spanning 120 km. The densities of these two intrusions lie between 2,720 and 3,300 kg/m³, with the maximum value in the upper center of the intrusions. Both intrusions have a bowl-like shape. Moreover, a significant magmatic body (Fig. 2d) characterized by a quasi-circular configuration is located adjacent to the west of the Marius-NH intrusion. This formation boasts a diameter of approximately 85 km and a thickness of about 6 km, positioned beneath the more subdued western inclines of the elevated Marius Hills plateau. A magmatic body of similar dimensions, encompassing a diameter of 80 km and a depth of 5 km, is identified beneath and in proximity to the Reiner crater (Fig. 2e). Both of these two magmatic bodies were neither discovered nor mentioned in previous studies[15, 24, 25, 26].

It is noteworthy that the inversion density model for study area A (Figs. 2c, 4) reveals a complex network of connectivity among the various volcanic systems under investigation. The Marius-NH and Marius-SH intrusions are interconnected with a narrow corridor of dense rocks extending approximately 80 km in length, 50 km in width, and 5 km in thickness (Figs. 2c, 4a, 4b, 4h). The narrow corridor is sill-like in structure and nearly horizontal. Furthermore, another two sill-like narrow corridors connect the two intrusions with the western border structures. The southern magmatic conduit, which traverses the intrusion beneath the Reiner crater, is approximately 140 km long, 50 km wide, and 5 km thick (Figs. 2c, 4a, 4b, 4e). The northern magmatic conduit, crossing the quasi-circular intrusion adjacent to the west side of the Marius-NH intrusion, measures about 120 km in length, 55 km in width, and 6 km in thickness (Figs. 2c, 4a, 4b, 4f).

To minimize the potential influence of crustal thinning, which is characterized by relatively long-wavelength positive gravity signals, we filtered signals with wavelengths under 11 degrees, and a subsequent 3D inversion was performed, detailed in Supplementary Figure 3. The inversion results (Supplementary Figs. 3c and 3d) indicate a modest reduction in the lateral spread of the PKT border structures and a 2-3 km shallowing of their bases. These two inversion models may more accurately reflect the border structures and remove the effect of crustal thinning. Nevertheless, we observed a reduction in the volume of the intrusions beneath Marius-SH and Marius-NH as well as the sill-like narrow corridors, where mantle uplift is deemed improbable[24, 25, 38, 39]. Moreover, the middle branch structures of the northwestern border become discontinuous. Thus, it seems that filtering out gravity signals lower than degree and order 11 has simultaneously affected the long-wavelength signals associated with shallow sources. As a result, we restrict our discussion to the inversion models (Figs. 2-5) derived from the Bouguer gravity data spanning from degree and order 7 to degree and order 660.

**Discussion**

Our research utilizing 3D density models, derived from the inversion of GRAIL Bouguer gravity data, has unveiled a complex magmatic architecture beneath the Procellarum Oceanus region. This intricate architecture reveals significant connectivity between the northern and southern magmatic intrusions beneath the MHVC and the PKT border magmatic structures. Our findings indicate an extensive network of magmatic connectivity beneath Oceanus Procellarum, which was previously unidentified and warrants further investigation and explanation.

The intrusion model[5] obtained through forward modeling of the average Bouguer gravity profile suggests that the middle branch in the northwest has a density contrast of 550 kg/m³, a width of 82 km (with a variability of -36 to +19 km), and a thickness of 6 km (with a range of -1 to +3 km). Our density model from 3D inversion (Figs. 3c, 5b, 5e) displays comparable parameters for the middle branch: it has a thickness of 5 to 7.5 km, a density ranging from ~2,720 to ~3,200 kg/m³ (~160 to ~640 kg/m³ for density contrasts), and spans approximately 80 km in width. Furthermore, the 3D inversion model shows that the middle branch is likely to have formed from three initially discrete magma chamber compartments, which gradually interconnected at higher levels throughout the filling process, a formation style reminiscent of the Great Dyke[40], where development also involved a sequence of initially isolated chambers. The eastern branch seems to have undergone a similar formation process; however, our inversion model only covers two of its chambers.

The magma evolution of the Marius Hills region was suggested to have extended over a prolonged period[19, 41, 42, 43], characterized by early edifice-forming and later mare-forming episodes. Recent research has mapped these mare flows on the plateau, with age estimates ranging from ~3.3 Ga to ~1 Ga [41, 44, 45, 46]. The Marius-SH and Marius-NH were suggested to be impact craters that have been buried by mare basalt and/or impact ejecta due to the circularity of the positive Bouguer gravity anomaly[15]. Other researchers[24, 25] indicated that extra high-density material is necessary in order to explain the large amplitude of the two positive gravity anomalies. The two main zones of high-density material were interpreted as the magma chambers that fed the overlying volcanism[24]. Some researchers suggested from forward modeling that a total thickness of ~10 km, with a density contrast of 350 kg/m³, is needed for a sill model[25]. However, the lack of significant topographic uplift at the Marius-SH area makes the existence of such large magma chambers unlikely. Consequently, an alternative dyke swarm model is favored, which could extend from the crust-mantle boundary to the floors of mare-filled impacts[25]. The

3D inversion density model (Figs. 2d, 2e) reveals that the magmatic intrusions beneath Marius-NH and Marius-SH extend from the surface/near surface to depths of 11.5 km and 12.5 km, with a thickness of ~10.5 km, and a density range of 2,720-3,300 kg/m$^3$ (an average density contrast of 315 kg/m$^3$). This is in good agreement with previously suggested dyke/sill models[24, 25]. In contrast, the 3D inversion model reveals more detailed characteristics of the intrusions and surrounding areas. The northern intrusion is situated beneath the highest topography, and many of the domes and cones within the Marius Hills plateau are positioned immediately above this intrusive body (Fig. 2d). The extent of the southern intrusion aligns with the overlying quasi-circular wrinkle ridges (Figs. 2e, Supplementary Fig. 2a), which primarily trend NNW-SSE with a relatively small number of volcanic vents situated on their broad arches[27]. The alignment between the wrinkle ridges and the southern intrusion likely reflects the local structural control of the wrinkle ridges by buried magmatic intrusion. The intrusion adjacent to the west of the Marius-NH intrusion (Figs. 2d, 4a, 4b, 4f) displays a quasi-circular shape in the horizontal plane. The corresponding Bouguer gravity anomaly (Fig. 2b) also displays a quasi-circular shape with a maximum value of 75 nT, which is characteristic of mare-filled impact craters[15, 26]. Hence, similar to the Marius-NH and Marius-SH anomalies, this anomaly might also represent a buried and filled impact crater. The placement of this intrusion coincides with the position of the groups of volcanic vents 5 and 7 as previously identified in the literature[27]. Given the spatial relationship between the volcanic vents and the intrusions, the three intrusions beneath the MHVC could have played a role in facilitating the volcanic eruptions occurring above them.

Dyke-related eruptions are the primary effusive style responsible for the extensive mare volcanism observed across the lunar maria[47]. Dyke propagation and emplacement lead to numerous distinct volcanic landforms[29]. For instance, when dykes reach the surface and erupt, they produce high-flux, high-volume effusive eruptions, resulting in extensive lava flows and sinuous rilles[26]. The dyke models beneath the Marius-SH and Marius-NH are suggested to extend to the crust-mantle boundary[25]. Our inversion model reveals a pair of sill-like intrusions with a thickness of ~10.5 km, which, however, do not extend to the crust-mantle boundary estimated at depths of 17-18 km for the study area[34]. The following factors could account for this discrepancy. Seismic data from the Apollo missions revealed a marked increase in velocity within the upper 20-25 km of the crust, correlating with porosity decrease due to increasing overburden pressure. For instance, it has been reported that lunar interior models M1 and M3 estimate subsurface densities of 2,680 kg/m³ at 10 km and 3,050 kg/m³ at 20 km for M1, and 2,760 kg/m³ at 10 km and 2,890 kg/m³ at 20 km for M3, respectively[48]. Our inversion model indicates that the density of the intrusions varies between 2,720 and 3,300 kg/m³, with the denser regions (2,900-3,300 kg/m³) predominantly at depths between 2-10 km (Figs. 4, 5). Given the inferred denser lunar crust at depth, it is plausible that the gravity anomaly from the deeper parts of the intrusions was much weaker due to the lower density contrast. Moreover, the amplitude of the gravity anomaly decays rapidly with depth[49]. As a result, the gravity anomaly caused by the intrusion models at depth was not detected and consequently failed to be recovered in the 3D density model.

Remarkably, the architecture illuminated by the 3D inversion models suggests a sophisticated and interlinked magmatic network beneath the Oceanus Procellarum (Figs. 2c, 4a, 4b, 4e, 4f, 4h). We interpret the three narrow near-horizontal sheeted magmatic structures identified as sill-like magmatic conduits. However, the origin of the magmatic conduits, their role in the regional magmatic transport, and their impact on the volcanic evolution in the Procellarum Oceanus region demand further inquiry. The phenomenon of magma traversing long distances within the Earth's crustal layers has been documented in a range of volcanic environments, including large igneous provinces[50, 51, 52], hot spot volcanoes[53, 54], mid-oceanic ridges[55], and island arc volcanoes[56, 57, 58]. It has been demonstrated that a network of sills and inclined sheets, known as a sill complex, enables magma to ascend vertically up to 12 km and spread laterally to distances of approximately 4,100 km[59, 60], indicating that volcanoes that are fed by these sill complexes can be situated a considerable distance laterally from their primary melt source. In contrast, it has been argued that the accelerated ascent of magma through dykes from deep lunar mantle sources generally favors vertical dykes over lateral sills[61]. However, in shallow crustal regions where the structures were altered by impacts, lateral formations can develop[61]. GRAIL gravity data analysis of about 1,200 complex craters reveals that the lunar megaregolith, which constitutes the uppermost and most fragmented layers of the lunar crust, is at least 8 km thick[36]. Lateral magmatic sill-like conduits identified within Oceanus Procellarum, spanning from the surface to a depth of 6-8 km correspond to this depth. Therefore, we inferred that the megaregolith's fragmented nature and significant thickness could have facilitated lateral magma transport in the shallow crustal layers. Higher-than-average heat flux[6] in the PKT region could also have played a role in these magmatic activities.

Previous studies have suggested that systemwide pressure weaknesses can propagate quickly through the magmatic web, facilitating lateral magma transport in different volcanic systems, such as beneath Hawaiian[54] and Kamchatka volcanoes[62]. The prolonged and extensive volcanic activity in the Marius Hills region, which produced domes, cones, and mare flows, might have caused pressure gradients between the plumbing system beneath the PKT border and the Marius Hills region. These pressure gradients could have facilitated rapid magma propagation through the fractures that prevail in lunar megaregolith from the PKT border to the Marius-SH and Marius-NH areas, as depicted in Figure. 6a. Furthermore, the pressure gradient caused by extensive eruptions at the northern Marius Hills might also have driven magma transportation from Marius-SH towards the Marius-NH area (Fig. 6b). This is consistent with the lack of topographic anomalies in the Marius-SH area and the smaller number of eruptive centers observed there. The magmatic conduit connecting Marius-SH and Marius-NH shows a NNW–SSE trend that aligns with the regional trend in the Marius Hills region, which is considered to be related to pre-existing weak zones/fractures[63, 64]. It is likely that the movement of magma between the Marius-SH and Marius-NH areas could proceed through the reactivation of these pre-existing fractures.

Lateral dyke intrusions traveling along the axis of magmatic rifts play an important role in accreting the uppermost part of the elastic–brittle oceanic lithosphere[65]. These intrusions are suggested to be driven by a horizontal pressure gradient along the rift, which is influenced by an increase in the thickness of the elastic-brittle layer[65] and/or a slope descending towards the ends of the rift segment[36]. The PKT border structures are interpreted as lava-filled rifts and underlying feeder dykes[5]. The formation of the rifts is considered to be a response to thermal stresses that arose from the differential cooling of the province compared to its surroundings. It is anticipated that horizontal pressure gradients might have developed during the formation of the rifts along the border of the PKT region, driving the lateral migration of magma along the slope of the rift zone. This process may have been further enhanced by pressure gradients caused by the volcanic eruptions at the Marius-NH and Marius-SH. The fractures resulting from the Reiner event (Fig. 6a) and the event associated with the formation of the inferred mare-filled crater (Fig. 4f), combined with the extensive pre-existing fractures in the lunar megaregolith, could have created channels that facilitate the flow of magma toward the Marius-SH and Marius-NH. This hypothesis aligns with the view that the lunar crust, extensively fractured and reworked by cratering, enables the exchange of substances and energy among different lunar systems[26]. Building on this framework, it is suggested that the magma plumbing system located beneath the western border of the PKT might not have achieved a state of complete cooling and solidification prior to these events, particularly before the formation of the Reiner crater, which dates back to the Eratosthenian epoch[66]. Consequently, the magma plumbing system beneath the PKT border is likely long-lasting. This is consistent with the surface ages of the nearby maria, which range from 1.2 to 4.0 Ga[67]. It has been suggested that the feeder dykes under the PKT border could have served as the magma plumbing system for much of the nearside mare volcanism[5].

The revelation of the connectivity between the PKT border structures and the Marius Hills intrusions from our inversion models might account for the compositional distinction between the domes and surrounding and embaying mare in the Marius region. According to the Moon Mineralogy Mapper ($M^3$) spectral analysis, the domes of the MHVC originated from more Fe/Mg-rich magmas, whereas the mare flows, which are considered younger than the domes[19, 42] were generated from more Ca-rich magmas[42]. Previous research suggests that these compositional differences could be indicative of the lengthy and intricate volcanic history of the region[19, 42, 43]. The connectivity revealed by the inversion model suggests that it is possible that the younger mare flows might originate from deep feeder dykes located beneath the western border of the PKT before the magma plumbing system completely cooled and solidified. Further detailed rheological and compositional analysis of the domes/cones and mare flows, along with surface age determination, will enhance our understanding of the origin of mare basalts in the MHVC. The 3D density model presented in this study suggests a possible origin through integrating MHVC into the broader context of the PKT.

Additionally, intrusions with various volumes have been discovered beneath many craters, including but not limited to the inferred filled and buried crater adjacent to the Marius-NH (Fig. 2d), Reiner crater (Fig. 2e), Marius crater (Fig. 2d), Briggs B crater (Fig. 3d), Lichtenberg crater (Fig. 3d), and Seleucus crater (Fig. 3e). A potential filled and buried impact crater (Fig. 4i) situated to the southwest of the Reiner crater has been detected. The corresponding Bouguer gravity anomaly (Fig. 2b) exhibits a quasi-circular shape with a maximum value of 100 nT. The intrusion beneath it directly connects to the PKT border structure (Fig. 4i). The smaller intrusion beneath the Marius crater (Figs. 2d, 4g) likely facilitated the floor-filling of the Marius crater. This process is believed to represent the most recent volcanic activity within the MHVC, triggered by the reactivation of fractures from the crater's impact event[19]. The intrusion might be connected to the sill-like conduit linking the Marius-NH and Marius-SH intrusions, however, due to the limited resolution of Bouguer gravity data and the inversion model, the interconnection is not clear. Previous research suggests that internal excess pressure generated fractures, impact-induced surface unloading, impact-induced fractures, and basin fractures played a significant role in facilitating magma ascent, intrusion, and eruption in mare-filled craters[29, 68, 69, 70, 71, 72, 73]. The observations from our 3D density models suggest that the accumulation of magma beneath craters may be quite common. The source of the magma does not necessarily originate directly from the deep lunar mantle. It is also possible that magma from within the lunar crust migrated through fractures to accumulate beneath impact craters. This insight helps clarify the dynamics of magma movement and storage beneath the lunar surface. Further improvements in lunar gravity data could lead to more accurate identification of smaller intrusions and the connections between different volcanic systems, which would significantly impact our understanding of lunar volcanic history and aid in the future exploration of lunar resources.

## Method

The 3D inversion technique applied in this research is based on the probabilistic method to generate 3D density models by inverting gravity data[74]. The method was assessed and proven successful in identifying the density structures of both the upper and lower crust of the Earth [9, 75, 76, 77, 78, 79, 80, 81]. A significant advantage of this inversion approach is its capacity to integrate a priori information regarding physical properties, spatial extent, depth, and orientation of the source body via the covariance matrix to construct an accurate and reliable solution.

The objective function employed for the inversion of gravity data is based on a probabilistic inversion method[74]:

$$\Phi = (\mathbf{Gm} - \mathbf{d}_{obs})^T \mathbf{C}_D^{-1} (\mathbf{Gm} - \mathbf{d}_{obs}) + (\mathbf{m} - \mathbf{m}_{apr})^T \mathbf{C}_M^{-1} (\mathbf{m} - \mathbf{m}_{apr}). \quad (1)$$

Here, $\mathbf{m}$ represents the model parameter vector, values of the densities of the cells, $\mathbf{d}$ is the gravity data vector, and $\mathbf{G}$ is the forward modeling matrix where $\mathbf{Gm} = \mathbf{d}$. The first term of equation (1) signifies the data misfit function, weighted by the data error covariance matrix, $\mathbf{C}_D$. The second term corresponds to the stabilizing function, which accounts for the deviation of model parameters $\mathbf{m}$ from a reference model $\mathbf{m}_{apr}$, weighted by the model covariance matrix $\mathbf{C}_M$. To minimize the objective function, the solution is derived below[74]

$$\widetilde{\mathbf{m}} = \mathbf{m}_{apr} + \mathbf{C}_M \mathbf{G}^T (\mathbf{G}\mathbf{C}_M \mathbf{G}^T + \mathbf{C}_D)^{-1} (\mathbf{d}_{obs} - \mathbf{G}\mathbf{m}_{apr}). \tag{2}$$

To counteract the natural decay of the forward modeling matrix with increasing depth, a depth weight function with the following form was used:

$$w(z) = 1/(z_{max} - z)^\beta, \tag{3}$$

where $z_{max}$ is the maximum depth of the inversion domain and $z$ is the depth of each cell. $\beta$ is the exponent of the depth weight function and typically ranges from 1 to 3. The value of $\beta$ influences the degree of density convergence along the depth direction. For instance, 3 is used in the Oasis Montaj software when inverting a flat-lying lens or sheet-type source. We have adopted $\beta=1$ for study area A and $\beta=2$ for study area B. These values were selected after comparing the inversion density models with those from previous studies through forward modeling[5, 24, 25]. The solution with the depth-weighting function is:

$$\widetilde{\mathbf{m}} = \mathbf{m}_{apr} + \mathbf{W}^{-1}\mathbf{C}_M \mathbf{W}^{-1}\mathbf{G}^T (\mathbf{G}\mathbf{W}^{-1}\mathbf{C}_M \mathbf{W}^{-1}\mathbf{G}^T + \mathbf{C}_D)^{-1} (\mathbf{d}_{obs} - \mathbf{G}\mathbf{m}_{apr}). \tag{4}$$

Before conducting the inversion, the Bouguer gravity data were transformed from spherical to Cartesian coordinates. The inversion domain was divided into numerous prism cells with dimensions of 1.5 km × 1.5 km × 0.25 km in the $x$, $y$, and depth directions, respectively. The depth range of the inversion domain was set at 20 km, with larger depth ranges of 30 and 40 km also tested, yielding nearly identical inversion results due to the influence of the chosen depth weighting function that can prevent the density from being erroneously pushed deeper as the depth of the inversion space increases. To mitigate edge effects during the inversion, the spatial dimensions of the inversion domain were expanded by 30 km in all lateral directions beyond the data coverage. This parameterization results in a model with 534× 526× 80 cells in study area A and 433× 467× 80 cells in study area B. The top of the inversion domain was constrained by the topography data. Therefore, the cells above the terrain were removed and not inverted. By constraining the top of the terrain, it is possible to mitigate the effects caused by inaccurate terrain-corrected densities and fully take into account the rapid changes in depth within the craters, thereby improving the accuracy of the inversion model. We observed positive topographic elevations in specific localized sections, mainly the rims of large craters located in the lower left corners of areas A (Fig. 2a) and B (Fig. 3a). During the inversion process, the topographic elevations of these regions were set to zero, as their very limited area did not impact our inversion results. A zero-reference model ($\mathbf{m}_{apr} = \mathbf{0}$) was used for the inversions, which corresponded to 2,560 kg/m³ in absolute density terms.

The parameters of the model covariance matrix $\mathbf{C}_M$ for areas A and B are listed in Table 1. For area A, we utilized a nested covariance model, as detailed in the second and third rows of Table 1, since the Bouguer gravity anomaly (Fig. 2b) within this region displays characteristics that span multiple scales. Specifically, the border structures of the PKT present larger linear formations trending NNW, while the anomalies at the Marius-SH and Marius-NH appear to be relatively smaller quasi-circular in nature. For area B, as detailed in the fourth row of Table 1, a non-nested covariance model was employed based on the observations of the Bouguer gravity anomaly (Fig. 3b), which displays narrow, linear features predominantly trending NNW.

Supplementary Figures. 4a and 4b show the difference maps between the observed gravity anomalies (Figs. 2a and 3a) and the predicted gravity anomalies calculated from the 3D inversion models (Figs. 4, 5). The difference maps show how well the inverted models resemble the observed gravity. The standard deviations of the differences for study areas A (Supplementary Fig. 4a) and B (Supplementary Fig. 4b) are 1.0 mGal and 1.1 mGal, respectively, with a mean of 0.0 mGal. In Supplementary Figure. 4a, we observe short wavelength fluctuations appearing as narrow strips, with widths of 6 to 10 km and lengths spanning 20 to 100 km, exhibiting a distinct S-N orientation. The short wavelength jitters presented in Supplementary Figure. 4b lack a clear directional pattern. The short wavelength jitters are mainly related to the orbit-parallel striping in the GRAIL data at high degree (Zhu et al., 2024). These fluctuations are not anticipated to disrupt the primary magmatic structures in this study, as the latter possess scales significantly larger than those of the strips.

The inversion method introduced above was proposed in the Cartesian coordinate system and can be directly extended to the spherical coordinate system. However, the inversion process can be much more efficient in the Cartesian coordinate system, where the inversion space can be divided into a series of adjacent prisms of the same size, and the FFT technique can highly speed up the inversion process[82]. To investigate the effect of the lunar curvature for this study, we inverted the Bouguer gravity in study areas A and B in the spherical coordinate system separately (Supplementary Figs. 5a, 5b). The resulting density models show a high degree of similarity to the models obtained in the Cartesian coordinate system (Figs. 2c, 3c) in terms of the density contrasts and the spatial extent of the inverted anomalies. However, due to the utilization of finer cells (1.5 km x 1.5 km x 0.25 km), the model in the Cartesian coordinate system reveals more detailed structures than in the spherical coordinate system (0.15° x 0.15° x 0.25 km), in which the cell size was limited by computational performance. Therefore, this study presents and discusses the inversion results using the Cartesian coordinate system.

# Figures

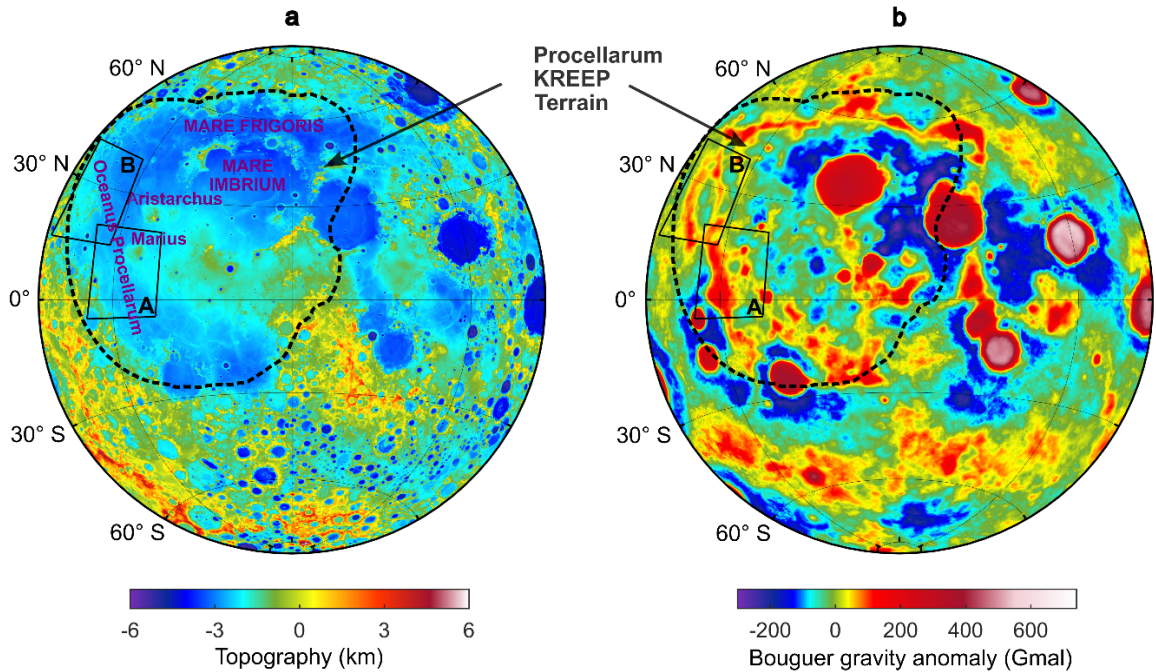

Fig. 1 | **Topographic and gravity features of the lunar nearside. a** Topography and **b** GRAIL gravity anomaly map shown with an azimuthal equal-area projection. The topography is derived from the MoonTopo2600p model[32] and the GRAIL gravity data were extracted from the GRGM1200B model[13]. Both the topography and gravity models are referenced to a lunar sphere radius of 1738 km. A crust density of 2,560 kg/m$^3$ was used for the Bouguer correction. Bouguer gravity data were filtered to include data between degree and order 7 and degree and order 660 to eliminate long-wavelength variations in the crustal structure and short-wavelength noise. The solid black polygons indicate the locations of two study areas labeled as A and B. Procellarum KREEP Terrane (PKT) boundary (black dashed line) is according to Jolliff et. al.[83] corresponding to the areas that incorporate most pixels that have >3.5 ppm Th.

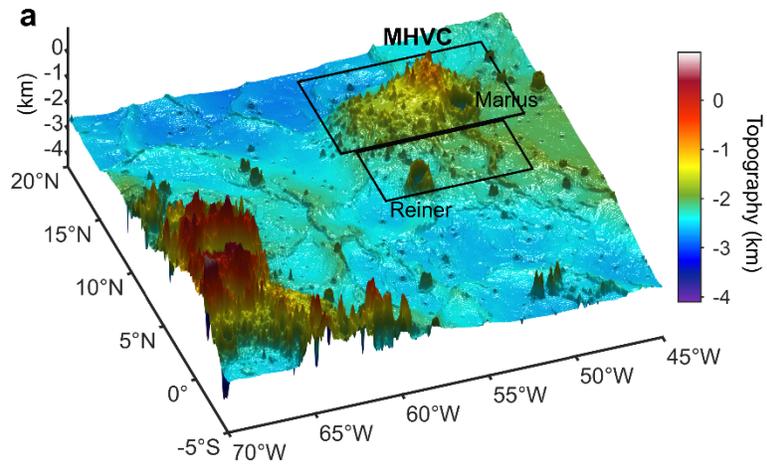

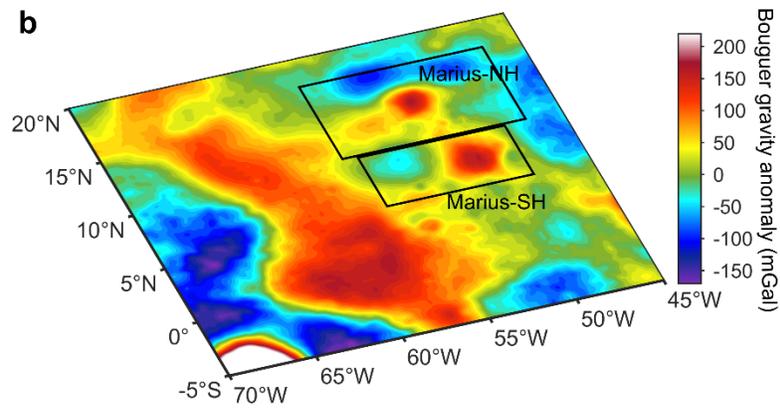

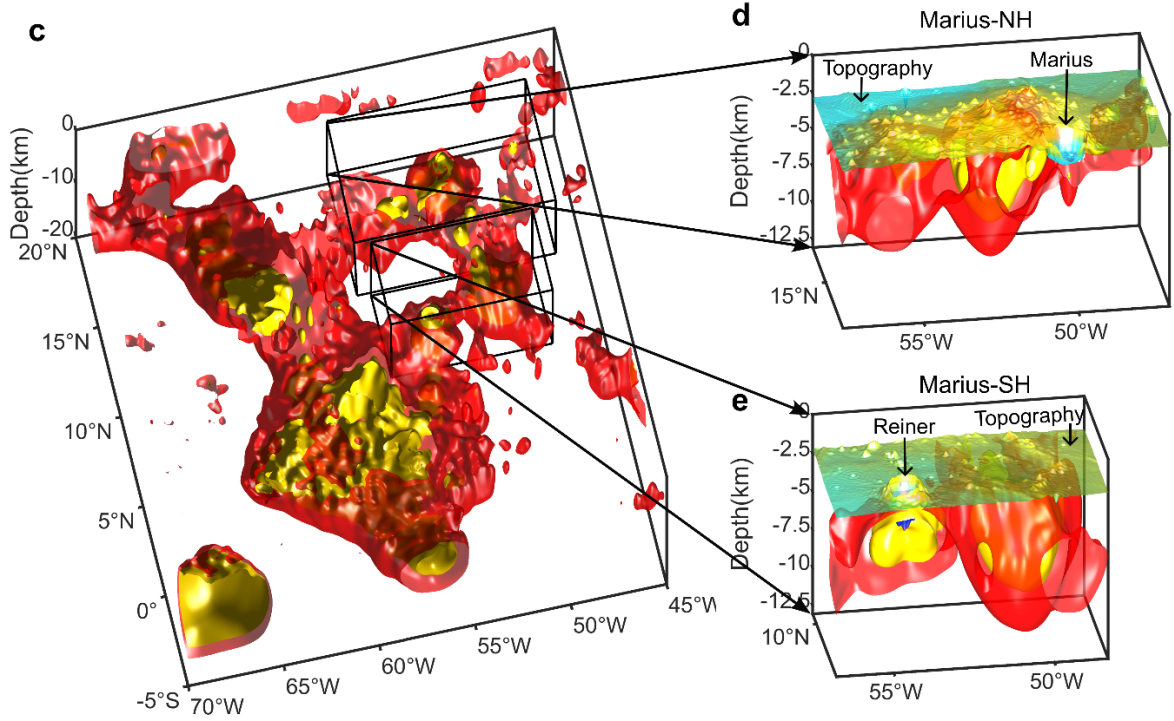

Fig. 2 | **Gravity anomaly features of study area A**. **a** Topography and **b** GRAIL Bouguer gravity anomaly map shown with an equidistant cylindrical projection. **c** Iso-surface map of the high-density anomalies from the 3D inversion. **d** Iso-surface map of the high-density anomalies at the northern MHVC. **e** Iso-surface map of the high-density anomalies at the southern MHVC. The location of study area A is indicated by solid black polygons in Fig. 1. The red surfaces in 2c-2e indicate the density iso-surface of 2,720 kg/m$^3$, which outlines high-density anomalies and is interpreted as magmatic bodies. The yellow surfaces depict a density iso-surface of 2,800 kg/m$^3$ to delineate the even denser magmatic bodies.

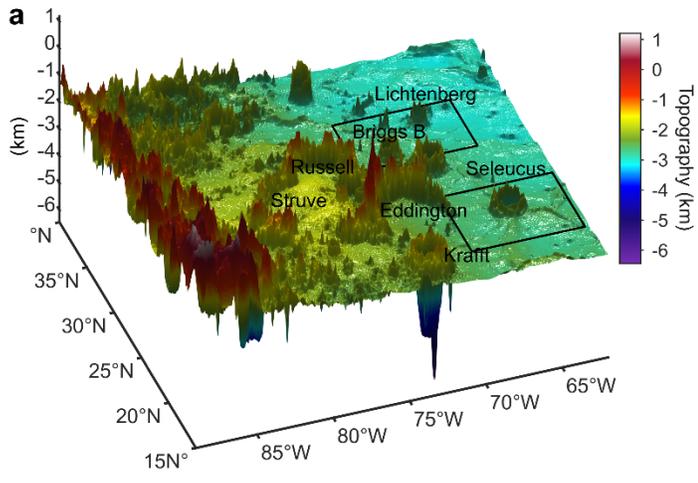
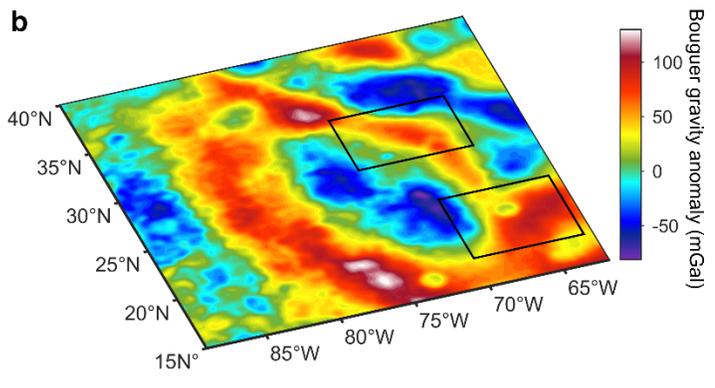
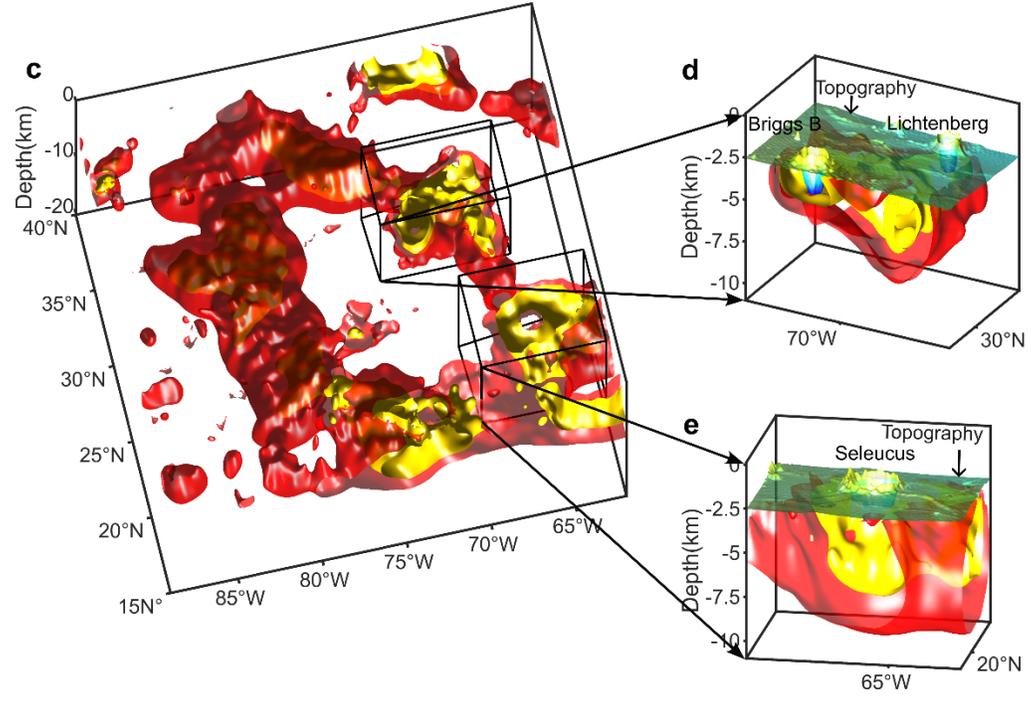

Fig. 3 | **Gravity anomaly features of study area B**. **a** Topography and **b** GRAIL Bouguer gravity anomaly map shown with an equidistant cylindrical projection. **c** Iso-surface map of the high-density anomalies from the 3D inversion. **d** Iso-surface map of the high-density anomalies beneath the Briggs B and Lichtenberg craters. **e** Iso-surface map of the high-density anomalies beneath the Seleucus crater. The location of study area B is indicated by solid black polygons in Fig. 1. The red surfaces in 3c-3e indicate the density iso-surface of 2,720 kg/m$^3$, which outlines high-density anomalies and is interpreted as magmatic bodies. The yellow surfaces depict a density iso-surface of 2,800 kg/m$^3$ to delineate the even denser magmatic bodies.

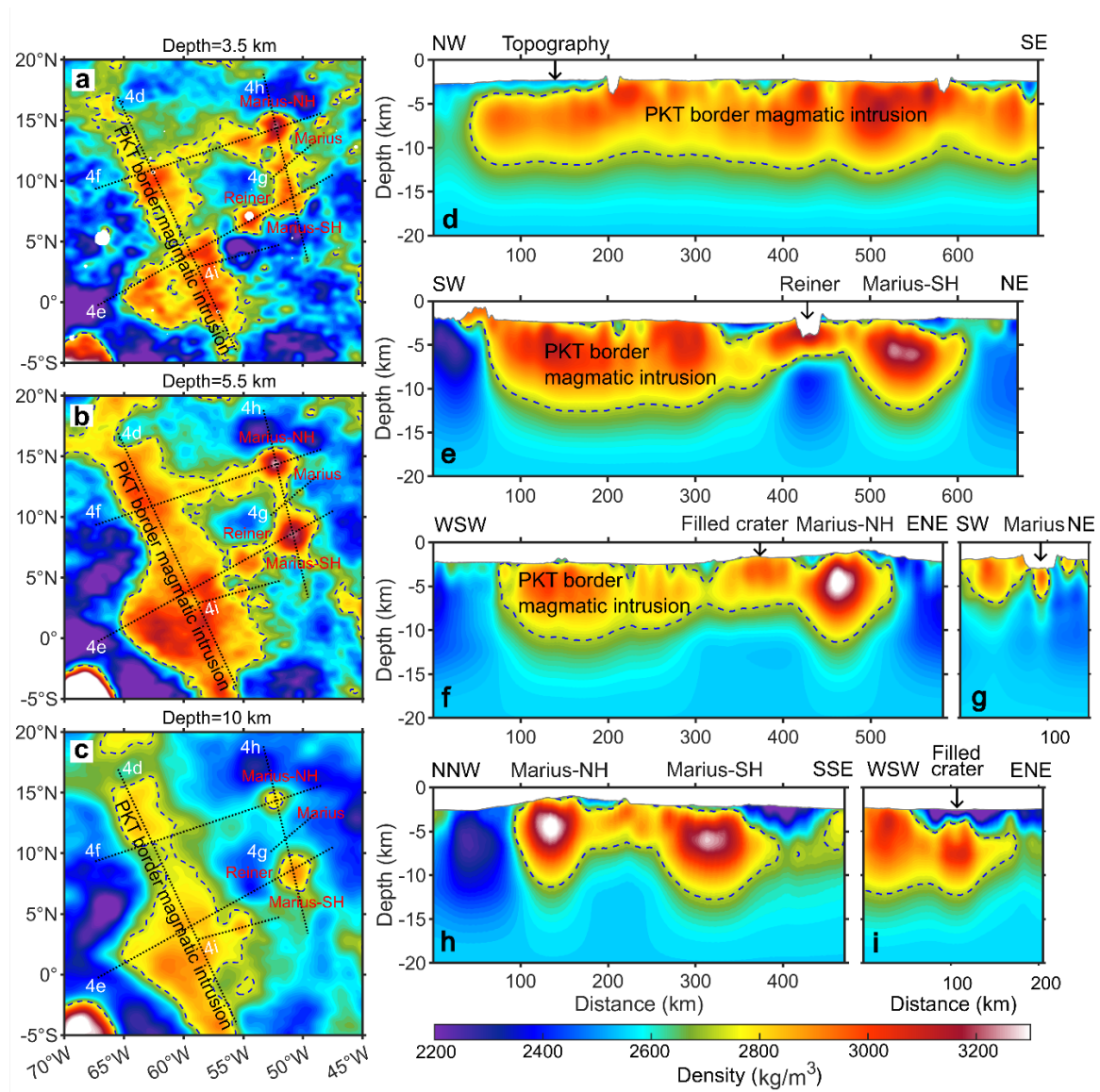

Fig. 4 | **Horizontal and vertical sections of the inverted 3D density model for study area A**. **a-c** Horizontal slices at depths of 3.5 km, 5.5 km, and 10 km. **d-i** Vertical depth slices. The locations of the vertical depth slices are indicated by black dotted lines on a-c. The blue dashed lines on a-i correlate with the 2,720 kg/m$^3$ contour line, which is taken to be the magmatic body surfaces. The vertical axis is exaggerated by a factor of 10 relative to the horizontal axis.

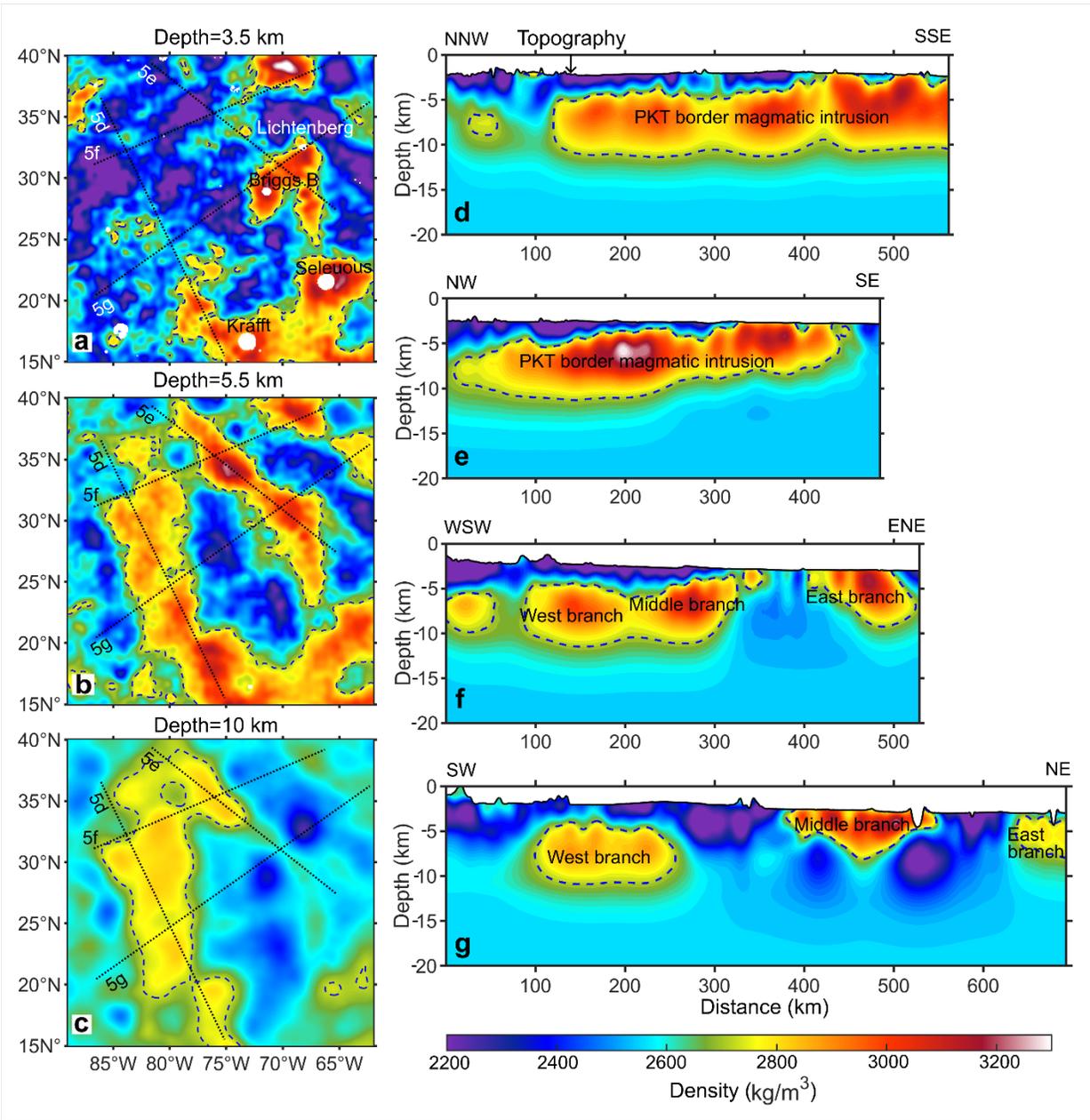

Fig. 5 | **Horizontal and vertical sections of the inverted 3D density model for study area B**. **a-c** Horizontal slices at depths of 3.5 km, 5.5 km, and 10 km. **d-g** Vertical depth slices. The locations of the vertical depth slices are indicated by black dotted lines on a-c. The blue dashed lines on a-i correlate with the 2,720 kg/m$^3$ contour line, which is taken to be the magmatic body surface. The vertical axis is exaggerated by a factor of 10 relative to the horizontal axis.

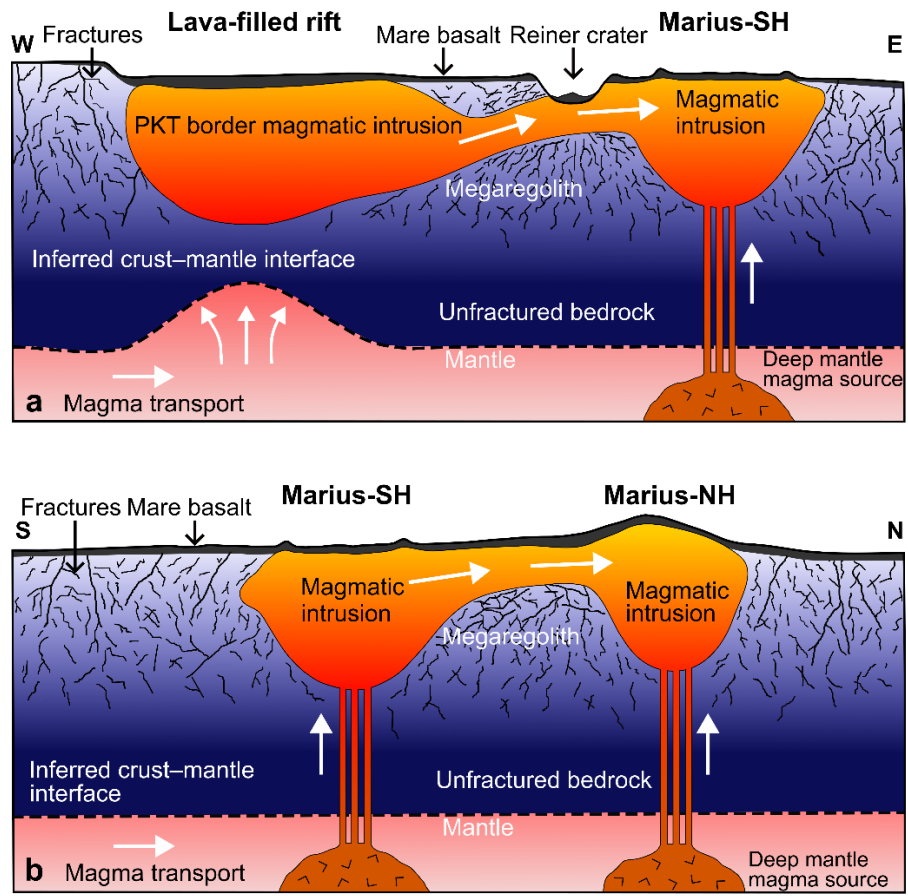

Fig. 6 | **Cartoon summarizing observations**. **a** The horizontal pressure gradient[65] along the rift, coupled with the pressure gradient caused by eruptions and intrusions at Marius-SH, drive the magma to propagate from the rift to Marius-SH. The fractures created by the Reiner impact event and the megaregolith's fragmented nature facilitate this magmatic activity. **b** Extensive eruptions and intrusions at Marius-NH cause pressure gradients to propagate rapidly through the regional weak zones/fractures[63, 64] to Marius-SH. Magma beneath Marius-SH is laterally transported along the fractures from Marius-SH to Marius-NH. To illustrate the magma transport process more clearly, the vertical axis is exaggerated by a factor of approximately 10 relative to the horizontal axis.

Table 1. The parameters of the model covariance matrix $\mathbf{C}_M$ for the areas A and B. A nested covariance model was used for the study area A, shown by S1 and S2. The direction corresponding to each range ($a_1$, $a_2$, and $a_3$) is indicated in the parentheses. $z$ represents the direction perpendicular to the ground. $C$ represents the value of sill; $C_0$ is the value of the nugget effect; Var_model indicates the variogram model used to calculate the covariance matrix[84].

|  |  | $a_1$/km | $a_2$/km | $a_3$/km | $C$/(kg/m)$^2$ | $C_0$ | Var_model |
|---|---|---|---|---|---|---|---|
| **Area A** | S1 | 80 (E) | 80 (N) | 10 (z) | 20000 | 0 | Gaussian |
|  | S2 | 800 (NNW) | 250 (ENE) | 10 (z) | 10000 | 0 | Gaussian |
| **Area B** |  | 600 (NW) | 200 (NE) | 10 (z) | 30000 | 0 | Gaussian |